# Polarization of light scattered by large aggregates

Ludmilla Kolokolova[a*], Daniel Mackowski[b]

[a] *University of Maryland, Department of Astronomy,, MD, 20742, USA,*

[b] *Auburn University, Mechanical Engineering Department, AL, 36849, USA*

* Corresponding author. tel. +1(301)405-1539, fax: +1(301)405-3559,

 Email: *ludmilla@astro.umd.edu* (L .Kolokolova)

Pages:   11

Figures:  7

Abstract

Study of cosmic dust and planetary aerosols indicate that some of them contain a large number of aggregates of the size that significantly exceeds the wavelengths of the visible light. In some cases such large aggregates may dominate in formation of the light scattering characteristics of the dust.  In this paper we present the results of computer modelling of light scattering by aggregates that contain more than 1000 monomers of submicron  size and study how their light scattering characteristics, specifically polarization, change with phase angle and  wavelength.  Such a modeling became possible due to development of a new version of MSTM (Multi Sphere T-Matrix) code for parallel computing. The results of the modeling are applied to the results of comet polarimetric observations to check if large aggregates dominate in formation of light scattering by comet dust.   We compare aggregates of different structure and porosity. We show that large aggregates of more than 98% porosity (e.g. ballistic cluster-cluster aggregates) have angular dependence of polarization almost identical to the Rayleigh one.  Large compact aggregates (less than 80% porosity) demonstrate the curves typical for solid particles.  This rules out too porous and too compact aggregates as typical comet dust particles.  We show that large aggregates not only can explain phase angle dependence of comet polarization in the near infrared but also may be responsible for the wavelength dependence of polarization, which can be related to their porosity.

# 1. Introduction

It is widely accepted now that many types of cosmic dust contain or are even primarily formed from aggregates of small particles. Examples are interstellar dust [1], interplanetary dust [2], comet dust [3], Titan [4] and some Earth [5] aerosols. In addition, many cosmic bodies (planetary satellites, asteroids) are covered by regolith that is a fluffy layer of aggregated particles. This made aggregates a focus of recent light-scattering modeling efforts, using both computational [6-8] and experimental [9, 10] methods, and stimulated development of the DDA [11] and MSTM [12] computational codes.

Until recently, DDA and MSTM predictions of aggregate scattering in the visible and IR were limited, for computational resource reasons, to aggregates having a characteristic size smaller than around 2 microns [6-8]. Such computations have described reasonably the visible wavelength observed properties, yet they also suggest that an improved description would be obtained from aggregates larger than those employed in the computations [13]. In addition, the larger aggregates better represented the observation data from comet dust in the thermal infrared [14]. Further evidence supporting the role of large aggregates was shown by the data obtained in the near infrared [15]. As one can see from Fig. 1, the comet polarization looks at the wavelength 2.2 μm practically the same way as it looks in the visible, at 0.6 μm, and it would therefore be reasonable to assume that the particles responsible for the polarization pattern are larger than 2.2 micron. An example of what happens in the case of smaller aggregates is presented in Fig. 2 where the phase angle (the angle Sun – object – observer, equal to 180° - scattering angle) dependence of linear polarization is shown for an aggregate of 256 constituent particles (monomers). The aggregate is the BPCA similar to the one that provided he best fit to the observations of comet dust in [13], i.e. the monomer size is 0.1μm and the composition was consistent with the composition of the comet Halley dust, containing silicates, amorphous carbon, and organics (for details see [6, 13]). One can see that the aggregate of 256 monomers rather well represents the solid curve shown in Fig. 1. However, at longer wavelengths the aggregate becomes too small and scatters light as a Rayleigh particle whereas the observational data still show negative polarization at small phase angles and rather low polarization maximum (Fig. 1) typical for particles comparable or larger than the wavelength.

In the following section we will check if considering the light scattering by large aggregates allows to solve this problem. Based on our earlier conclusions [17] about the wavelength dependence of polarization in the near infrared on the porosity of aggregates, we will also check how different porosity of aggregates affects their behaviour in the near infrared (Section 3). We will use comet dust as a test object and will compare our results with the cometary polarimetric data.

## 2. Computer modeling of large aggregates of spherical particles

To make the scattering particles larger than Rayleigh one in the near infrared, we built large ballistic aggregates of various porosity. We also modeled a particle as a spherical volume medium filled with randomly distributed spheres (similar to the one used in [18]) to see if it provides the results similar to those for ballistic aggregates of the same size and porosity. To make the results applicable to comets, we used the model of monomers described in Section 1, i.e., the radius of monomers r=0.1 µm and the material has Halley-like composition. This not only provides the best fit to the observational data but is also consistent with the evolutionary ideas of the comet dust formation [19] and *in situ* data for comet Halley [20]. To separate the influence of porosity from the influence of the optical constants, we fixed the refractive index at its value for the wavelength 0.45 µm, i.e. m= 1.88-i0.47 [13]. We considered aggregates of 1024 monomers; larger particles appeared to be out of the capability of our computers. To calculate the characteristic radius of the aggregates we followed the approach described in [14], namely, $r_C = \sqrt{(5/3)}\, r_g$ where $r_g$ denotes the gyration radius of the aggregate; this defines the porosity of the aggregates as $P = 1-N\, (r/r_C)^3$ where $r$ is the radius of monomers and N their number.
Their characteristic radius appeared to be around 4-5 µm if we built the aggregates as Ballistic Cluster Cluster Aggregates (BCCA) and ≈2 µm for BPCA (Ballistic Particle-Cluster Aggregate). This made the size parameters of the aggregates significantly larger than unity even for the wavelength 4.4 µm.

Light-scattering modeling of particles of this size became possible after developing a version of the Multiple Sphere T-Matrix (MSTM) code for parallel computing described in [21]. For our computations we used the compute cluster *yorp* of the Astronomy Department at the University of Maryland that allowed us to use 2 CPU of 6 cores each represented by Intel Xeon X5670 2.9 GHz processors with the total RAM equal to 24 GB.

We started with testing BPCA aggregates of the porosity 80% and BCCA aggregates of the porosity 98%. These porosities are close to the low limit of porosities for large ballistic aggregates of corresponding types. The results for the wavelength 2.2 μm are shown in Fig. 3. One can see that the BPCA, a rather compact aggregate, demonstrates a resonance structure typical for solid particles of this size. It seems likely that since such an aggregate become smaller than the wavelengths in the near-infrared, and its monomers are very closely located, the light interacts with the aggregate as a whole and loses sensitivity to its structure. Of course, considering a polidisperse mixture of such aggregates we could smooth out the resonance structure. However, the numerous previous attempts to model comet dust as an ensemble of solid polidisperse and even multishaped particles (see [22]) failed to reproduce all photopolarimetric properties of the comet dust. Thus, the particles that interact with the light as solid ones are not good candidates for particles that dominate light scattering properties of comet dust. On the other hand, very porous particles, as the considered BCCA, interact with the light as a cloud of rather distant monomers and, thus, properties of such aggregates are mainly defined by the properties of individual monomers, which in our case are much smaller than the wavelength. In the result the BCCA particle demonstrate the light-scattering behaviour typical for Rayleigh particles. Since we do not see any of these behaviours in the comet data, we need to assume that the comet dust particles have the porosity higher than 80% and lower than 98%. This immediately rules out BCCA as particles that dominate in comet dust. Rather porous BPCA are still possible, however, large BPCA usually have porosity not more than 85-86%. A good candidate for aggregates of porosities within 85-95% could be rebound aggregates [23], i.e. aggregates produced at collisions of BCCAs.

Fig.4

For a further application of our computer modeling to the comet polarimetric data we checked if large aggregates can reproduce the observed wavelength dependence of polarization. These observations are described in detail by Kiselev et al [24], and are characterized by increase of polarization in the visible and decrease in the near infrared as shown in Fig. 4. Kiselev et al also showed that for some comets the polarization can decrease even in the visible as it typically does for asteroids and interplanetary dust. The wavelength behavior of comet polarization can be explained based on the idea of electromagnetic interaction (coupling) between the monomers in aggregates. It was shown in [17] that coupling between monomers is significant only when the size and spacing of the monomers is less than the wavelength. Such conditions lead to a depolarization of the scattered radiation, and the degree of depolarization becomes stronger as the number of monomers covered by a single wavelength increases (see [17] for details). From this, it was concluded that (1) compact aggregates should depolarize the light stronger than porous aggregates, and (2) depolarization from aggregates having characteristic sizes larger than the wavelength should increase with increasing wavelength.

Fig. 5

Figure 5 shows the results of our computations for two BPCA particles consisting of 1024 monomers. One of the aggregates has the porosity 90% and the other one is of 85% porosity. The figure shows their maximum polarization, which is usually reached at the phase angles around 85-105°. For both aggregates the maximum polarization increases with wavelength as the wavelengths is shorter than 3 µm. At longer wavelengths, the increase slows down for the more porous aggregate and changes to a decrease for the more compact aggregate reminding the trend we see in Fig. 4. This result is consistent with the decrease in polarization of the scattered light caused by electromagnetic interaction described in [17]. However, at wavelengths longer than 4 µm, we see an increase of polarization again. Most likely, this is because the wavelength becomes larger than the aggregate size. In this case, the interaction of the electromagnetic wave with the aggregate is mostly affected by the aggregate as a whole approaching and finally reaching the Rayleigh regime; see more evidence of this in Fig. 6. Of course, calculations involving larger aggregates (with larger number of monomers) would need to be performed to provide further evidence. The observations at longer wavelengths would be also helpful, particularly in order to specify the size of dominating dust particles. However, at the heliocentric distances where comets reach phase angles around 90° polarimetric data for the wavelengths longer than 4 μm become unreliable as they are highly affected by thermal emission. The decrease in polarization seen in Fig. 5 starts later than it is observed in comets (cf. Fig. 4). However, as it was shown in [17], not only porosity but also the refractive index of the material affects the electromagnetic interaction. As we said above, in this study we fixed the refractive index at m=1.88-i0.47 to isolate the effects of porosity and optical constants. It is very likely that the realistic spectral change in the refractive index would shift the wavelength where the depolarization becomes efficient to shorter wavelengths.

Apparently, the influence of aggregate characteristics on the wavelength dependence of polarization requires a careful survey. However, it is already clear that it can be a useful tool for understanding aggregate properties, specifically their porosity. We note that such calculations are not cheap to perform. In the case of the yorp computer cluster described above, the calculations for the considered aggregates of 1024 monomers required CPU time about 30 hours for the wavelength 0.6 μm, 10 hours for 1.1 μm, although it dropped to 0.1 hours for 4.4 μm.

Fig.6

We also tested how important is the fact that the monomers in aggregates are connected to each other. For this, we modeled a particle with characteristics similar the BPCA described in the previous paragraph, i.e., 85% porosity and identical number and type of monomers, yet with the monomers randomly distributed within a spherical volume. Such types of "particles" have been used to model light scattering characteristics of planetary regolith [18, 25, 26] and even aggregated particles [27] as this model allows to produce "particles" of controlled porosity. Some of the monomers in such models touch each other, but many of them are not connected to any other monomer. The results for this and equal-number and porosity BPCA are shown in Fig. 6. One can see that at short wavelengths both particles demonstrate similar behavior, although the light scattered by a real BPCA aggregate is more depolarized, probably, indicating stronger interaction between the particles if they are connected. However, at longer wavelengths BPCA still shows a smooth curve whereas the spherical volume starts showing some ripples that at longer wavelengths convert to a noticeable resonant structure typical for solid particles. At longer wavelengths both aggregates behave as Rayleigh particles. Explanation of the difference in the behavior of the two particles may be that random distribution of monomers in the volume, absence of a fractal structure, makes the volume working as an "effective medium", where the refractive index of the medium can be calculated by averaging (using Bruggeman, Maxwell Garnett or another mixing rules) the refractive indices of the components and then the medium acts as a solid sphere with the corresponding effective refractive index. Notice that changing the shape of the volume to an ellipsoid or any other non-spherical randomly oriented particle smoothes out the ripples. However, large BPCAs approach spherical shape and still do not show any resonant structure in the behavior of their light-scattering characteristics. This result is a good illustration of how important for light scattering characteristics is the fact that the monomers are in contact with each other. It is a warning that modeling light scattering by realistic aggregates one need to build them properly, checking if each particle connects with another one at least once.

Fig.5

## 3. Conclusions

Our computer modeling shows that to reproduce the near infrared polarimetric data for comets we need to consider aggregates larger than the wavelength. Thus, it provides evidence that light scattering by comet dust is dominated by large aggregates. This is not surprising as domination of larger aggregates was one of the conclusions of the Stardust sample return experiment. For example, Flynn [28] states "The Wild 2 particles collected by the Stardust spacecraft were generally weakly bound aggregates of submicron grains" and shows in his Figure 2 that the main mass of the collected samples came from large particles. The domination of large particles was also concluded from thermal infrared and in-situ Halley measurements as described in [14]. Specifically, [14] showed that size distribution of the dust particles in the coma of comet Halley [29] results in the total dust cross section dominated by particles larger than 1 μm. Fig.7 confirms the dominating role of large particles. It shows that for an ensemble of BPCA with Halley-like mass distribution the contribution of particles in the scattered light becomes larger if the number of monomers increases. This results from an increase of both scattering efficiency and cross section of BPCAs whose combined effect overpowers the decrease of the number of particles in the power-law size distribution with the power ≤2.75 that comet Halley had in the considered range of sizes [30].

Fig.7

One more conclusion of this study is that considering large aggregates in a broad range of wavelengths we can study how porosity affects their polarization. Specifically, the change of the polarimetric color (the spectral gradient of polarization) from positive to negative is strongly affected by the porosity and can be a good tool to determine its value.

Finally, our computations show that a proper model of the aggregate is important to correctly reproduce the light scattering characteristics of realistic particles.

**Acknowledgement**

The work was accomplished thanks to the grant of NASA Exobiology and Astrobiology Program. We also acknowledge Lev Nagdimunov who developed programs for building and displaying BCCA and BPCA particles. We also thankful to the anonymous referee whose comments allowed us to reveal more interesting results in our computational data.

Figure captions

Fig. 1. Phase angle dependence of polarization of comets in the visible (red filter, shown by solid line) and near infrared. The NIR data are for the wavelength 2.2 µm, except the points for comets Lulin and Gunn at small phase angles and for comet SW3 that were obtained at 1.65 µm. The picture is adapted from [16].

Fig. 2. Computed phase angle dependence of polarization for an aggregate of 256 submicron monomers at the wavelength 0.45 µm (solid line) and 2.2 µm (dotted line). The aggregate is shown in the insert. Its size parameter is 8.8 at 0.45 µm and 1.8 at 2.2 µm.

Fig. 3. Phase angle dependence of polarization for aggregates of 1024 monomers. The computations were done for the wavelength 2.2 µm. Left panel shows the results for aggregate of porosity 80%, the right panel shows the results for the porosity 98%. The model aggregates are shown in inserts.

Fig. 4. Wavelength dependence of polarization for comet Hale-Bopp. Notice increase of polarization in the visible and decrease in the near infrared. Similar behavior is typical for other comets (see [24]). The figure is adapted from [24].

Fig. 5. Wavelength dependence of polarization for BPCA aggregates of porosity 85% (o) and 90% (+). Notice the difference in the wavelength dependence of polarization for NIR wavelengths and the tendency to approach the Rayleigh regime at longer wavelengths for both particles.

Fig. 6. Phase angle dependence of polarization for a BPCA of 1024 monomers (top panel) and for a particle built as a spherical volume filled with the same monomers (bottom panel). Both aggregates are of porosity 85%; they are shown in inserts. The wavelengths are (left to right) 0.6 µm, 1.1 µm, 2.2 µm, 4.4 µm, and 6 µm. Notice Rayleigh behavior of both "particles" at longer wavelengths. Characteristic radius of both aggregates is 1.9 µm.

Fig. 7 Fractional scattering cross section per mass decade (following [14]) for BPCA with the size distribution measured *in situ* for comet Halley [28] (left axes, solid line). The dotted line shows the scattering efficiency, $Q_{sca}$, for the corresponding BPCA (right axes), the dashed line shows its effective radius, $a_{eff}$.

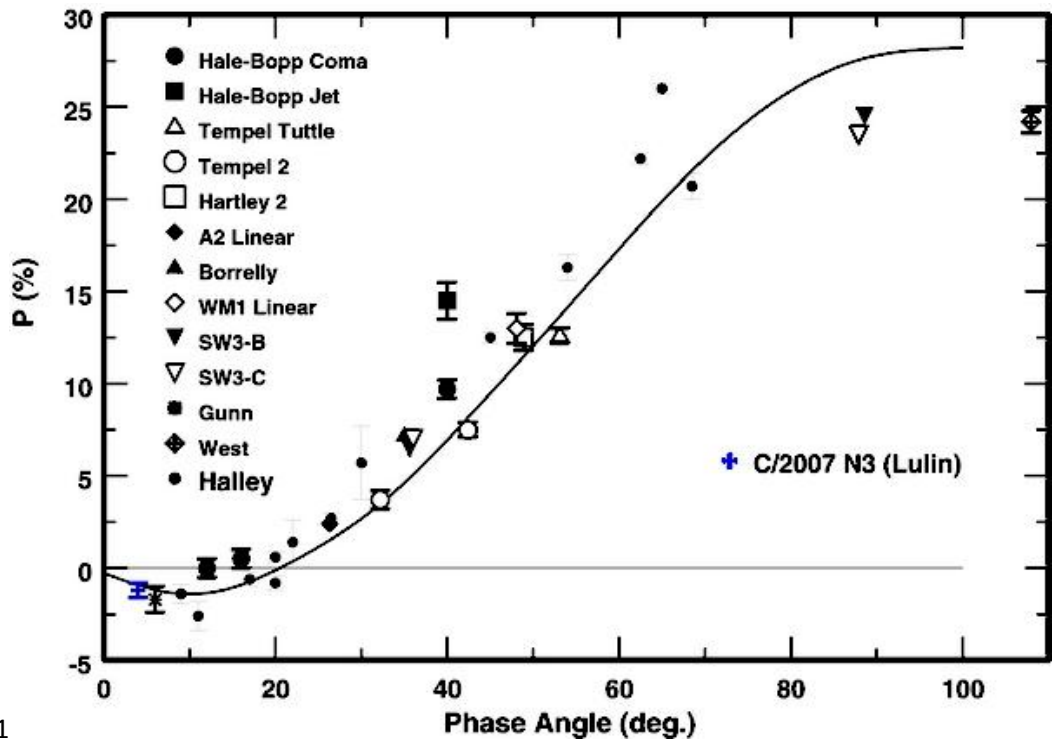

Fig. 1

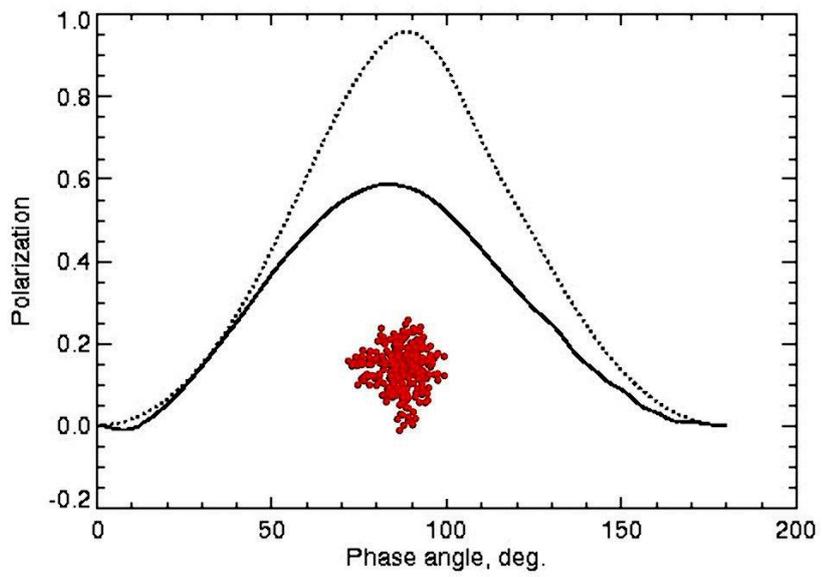

Fig. 2

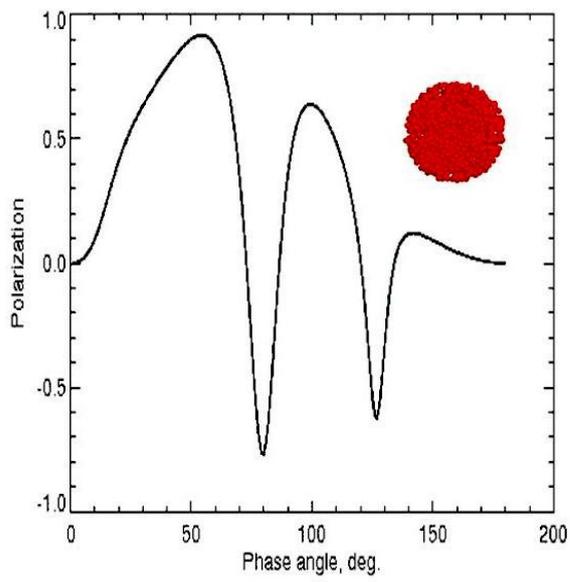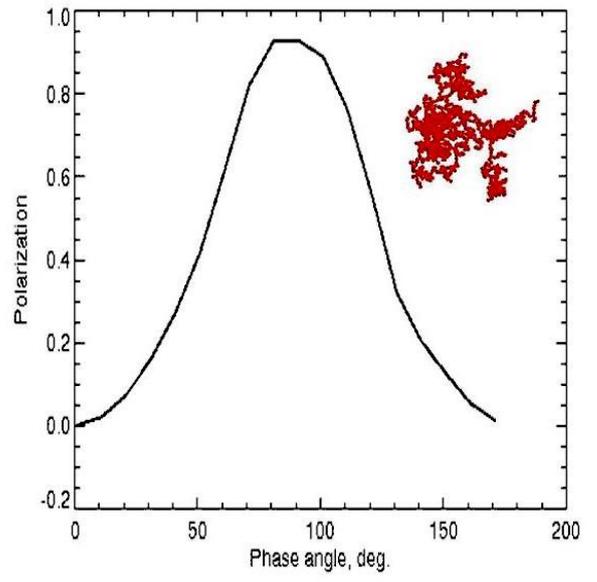

Fig. 3

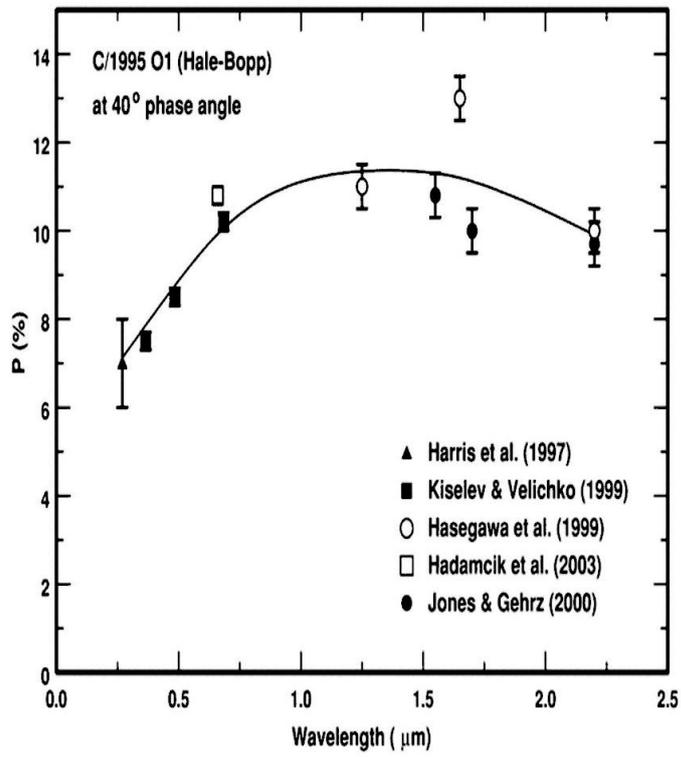

Fig. 4

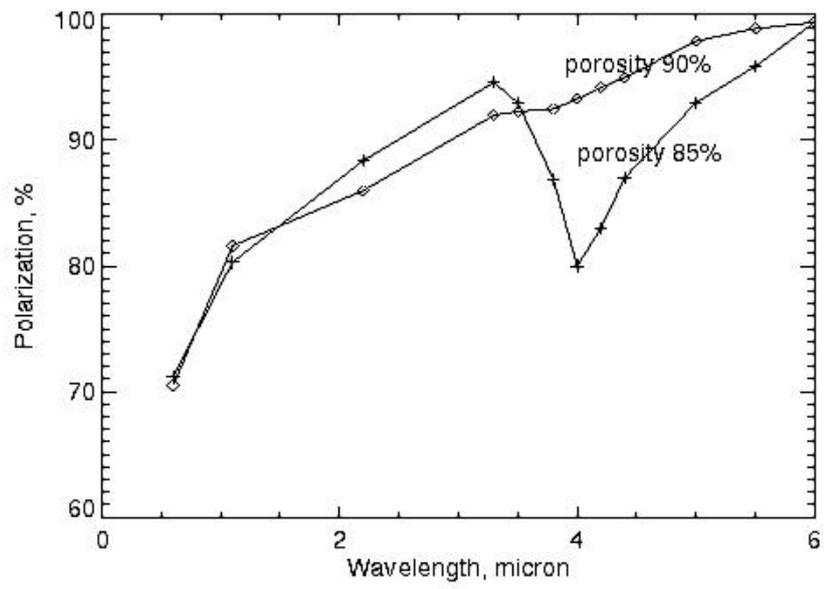

Fig. 5

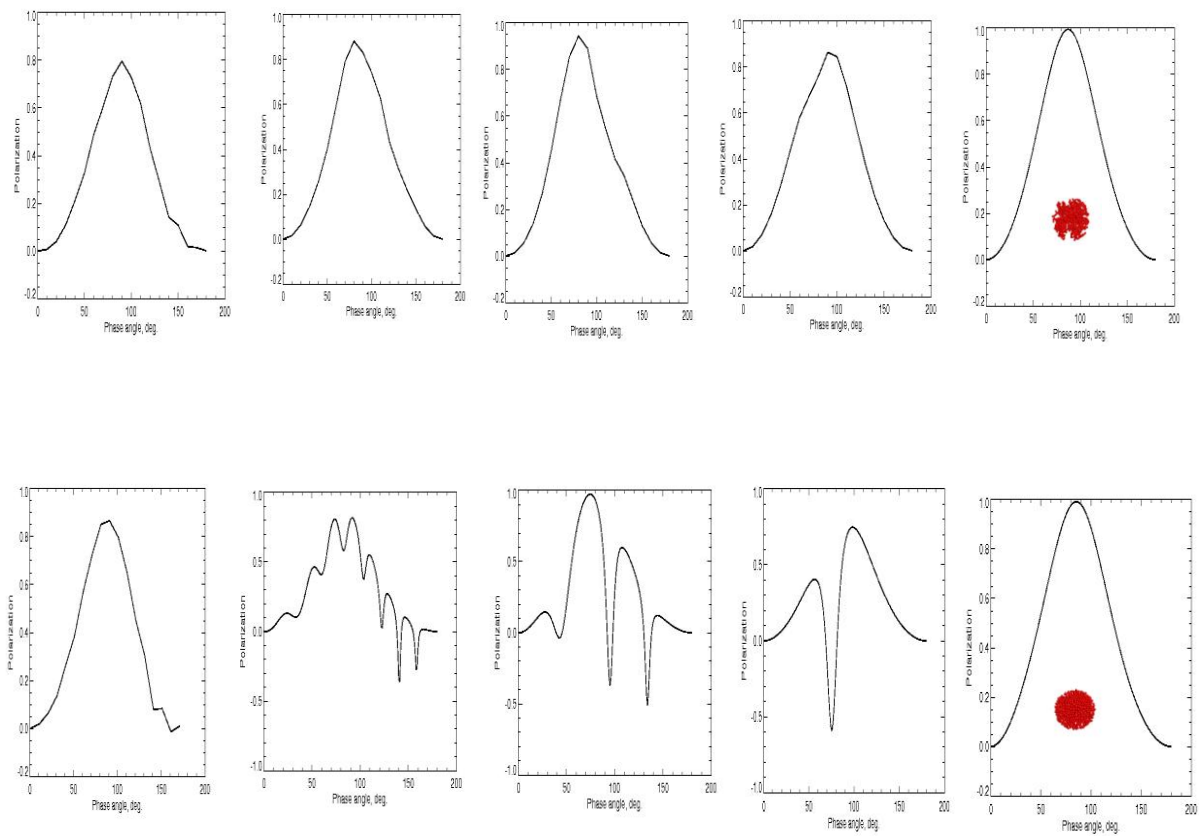

Fig. 6

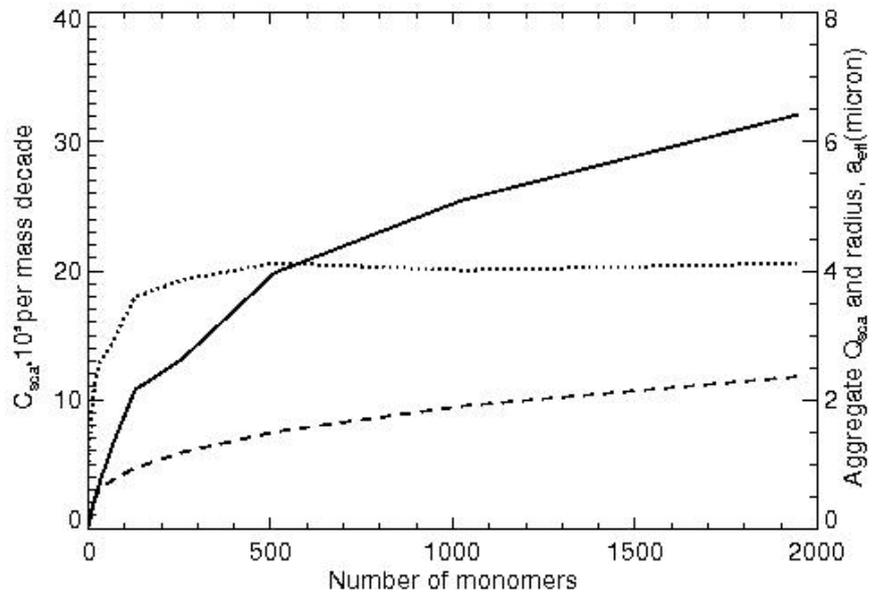

Fig. 7